\documentclass[reprint,aps,prl,showpacs,twocolumn]{revtex4-1}

\usepackage[utf8]{inputenc}
\usepackage{mathptmx}
\usepackage{mathtools}
\usepackage{amsmath} 
\usepackage{braket}
\usepackage{float}
\DeclarePairedDelimiter\abs{\lvert}{\rvert}%
\usepackage{lipsum}  
\usepackage{svg}

\usepackage{mathptmx}
\usepackage{stackrel}
\usepackage{mathtools}
\usepackage{amsmath} 
\usepackage{braket}
\usepackage{float}
\usepackage{lipsum}  
\usepackage{svg}

\usepackage[sort&compress]{natbib}
\usepackage{mhchem}
\usepackage{changepage}
\usepackage{float}
\usepackage{mathrsfs}
\usepackage{tikz}
\usepackage{mathrsfs}
\usepackage{tikz}

\date{}

\begin{document}

\title{A Unified Approach to Gated Reactions on Networks}

\author{{\normalsize{}Yuval Scher$^{1,}$}
{\normalsize{}}}
\email{yuvalscher@mail.tau.ac.il}

\author{{\normalsize{}Shlomi Reuveni$^{1,}$}
{\normalsize{}}}
\email{shlomire@tauex.tau.ac.il}

\affiliation{\noindent \textit{$^{1}$School of Chemistry, Center for the Physics \& Chemistry of Living Systems, Ratner Institute for Single Molecule Chemistry, and the Sackler Center for Computational Molecular \& Materials Science, Tel Aviv University, 6997801, Tel Aviv, Israel}}

\date{\today}


\begin{abstract}
\noindent For two molecules to react they first have to meet. Yet, reaction times are rarely on par with the first-passage times that govern  such molecular encounters. A prime reason for this discrepancy is stochastic transitions between reactive and non-reactive molecular states, which results in effective gating of product formation and altered reaction kinetics. To better understand this phenomenon we develop a unifying approach to gated reactions on networks. We first show that the mean and distribution of the gated reaction time can always be expressed in terms of  ungated first-passage and return times. This relation between gated and ungated kinetics is then explored to reveal universal features of gated reactions. The latter are exemplified using a diverse set of case studies which are also used to expose the exotic kinetics that arises due to molecular gating.  
\end{abstract}

\maketitle

The first-passage properties of a random walker are central to the study and analysis of stochastic phenomena \cite{redner2001guide,Bray2013review,metzler2014first,klafter2011first}. The classic problem of a random walker in search of a target arises naturally in a variety of fields, be it biology \cite{iyer2016first,Chou2014first,Polizzi2016first,alberty1958application,szabo1978kinetics,karplus1979diffusion,boehr2009role, changeux2011conformational,vogt2012conformational}, finance \cite{steele2012stochastic,Chicheportiche2014stochastic} or network science \cite{Sood2004first,Condamin2007first,Tejedor2009first,Reuveni2012first,hwang2012first,Tishby2017first}, to name a few. Markedly, the first-passage problem is central to the  theory of diffusion limited reactions, where a reaction between two species can be modeled as stochastic motion that terminates on contact \cite{von1916drei, smoluchowski1918versuch,benichou2010geometry}. However, while having two molecules at the same place and at the same time is a necessary condition for the occurrence of a chemical reaction, more is oftentimes required for the reaction to actually take place. 

It has long been realized that in order to better depict chemical reactions one has to consider the possibility of infertile molecular collisions \cite{collins1949diffusion,berezhkovskii1989kinetics,frisch1952diffusional,sano1979partially,shoup1981diffusion,szabo1984localized,szabo1989theory,burlatsky1990diffusion,kim1992theory,calef1983diffusion,weiss1986overview}. This realization later matured to the concept of gated reactions, which occur only when molecules collide while in a reactive state (alternatively, the gate is open) \cite{mccammon1981gated,szabo1980first,berezhkovskii1997smoluchowski,szabo1982stochastically,northrup1982rate,budde1995transient,caceres1995theory,re1996survival,spouge1996single,zhou1996theory,sheu1997survival,makhnovskii1998stochastic,sheu1999first,bandyopadhyay2000theoretical,benichou2000kinetics,godec2017first}, see Fig. 1. A similar idea of a gated boundary was applied in narrow gated-escape problems \cite{reingruber2009gated,bressloff2015escape} and search for gated targets \cite{shin2018molecular,mercado2019first}. Depending on the context, the first-time of arriving at the boundary while at the reactive state is interchangeably referred to as reaction time, first-absorption time or first-hitting time. Throughout the years many examples of gated processes were explicitly solved for, but a general understanding of gated-reaction kinetics is still missing. A notable step in that direction is the formalism developed by  Spouge, Szabo and Weiss \cite{spouge1996single}. 

At the heart of this letter is a renewal approach that is used to build a unified framework to gated reactions on networks. Specifically, by employing this renewal approach we provide for simple and general relations between gated and ungated reaction times. These, in turn, are used to show that it is enough to solve for the ungated problem to readily obtain a solution for the corresponding gated problem. Solving the ungated problem is generally a much simpler task, which e.g., allows for clean re-derivation of classic results that were previously obtained via brute-force methods. Moreover, the relations derived below open the door for systematic and widespread analysis of gated kinetics by providing ready-made solutions in all cases where the underlying, i.e., ungated, problem has already been (or can be) solved. This important feature of our framework has already proven  extremely useful in the context of stochastic resetting where similar renewal methods were heavily employed \cite{renewal1,renewal2,renewal3,renewal4,renewal5,renewal6,renewal7,renewal8,renewal9,renewal10,renewal11,renewal12,renewal13,renewal14,renewal15,renewal16,renewal17}.

\begin{figure}[t]
\begin{centering}
\includegraphics[width=0.75\linewidth]{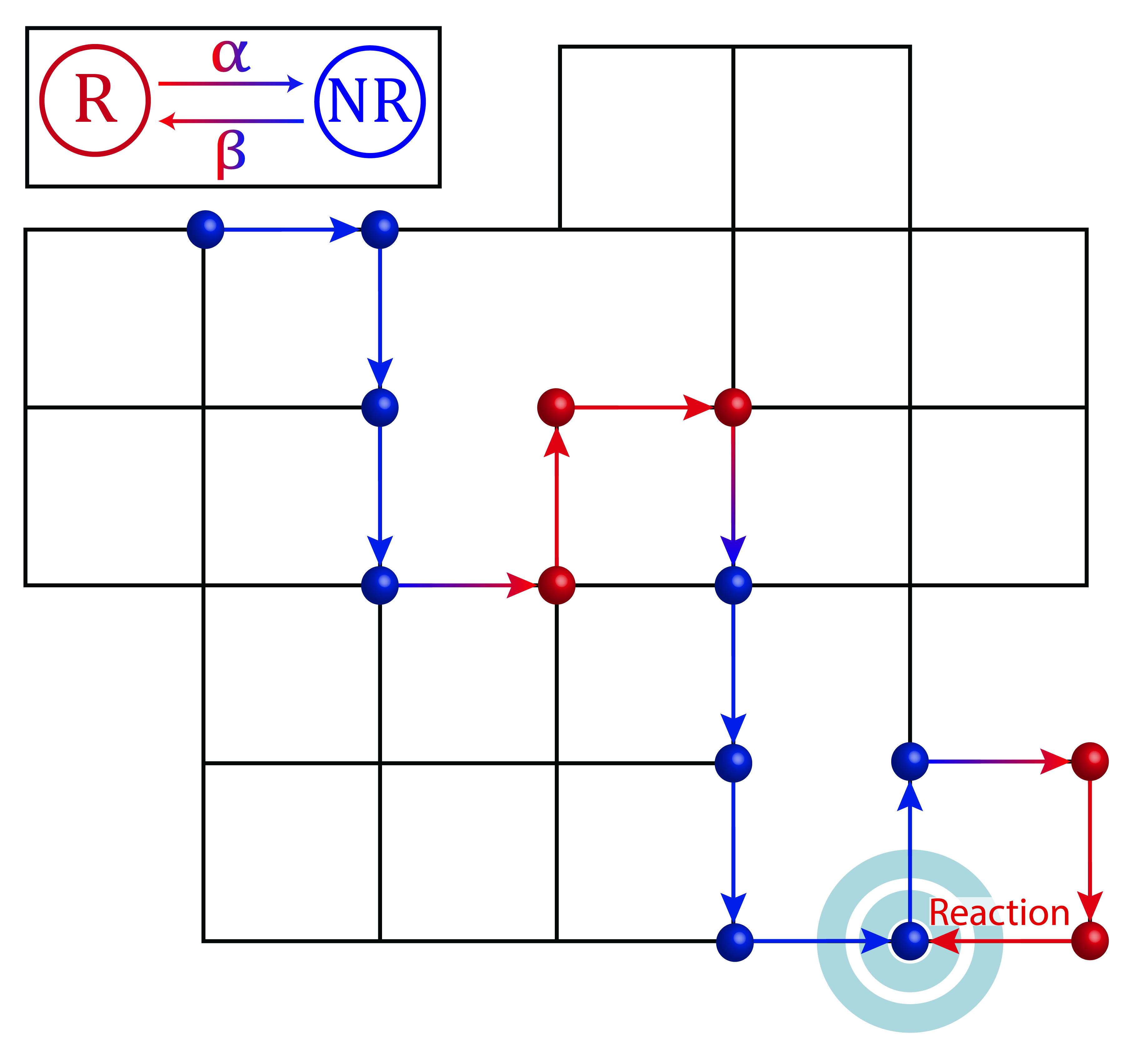}
\caption{Illustration of a gated reaction on a network. The reactant, here depicted as a small sphere, performs a random walk while switching between the non-reactive (NR, blue) and reactive (R, red) states. Reaction occurs when two conditions are met: (i) the reactant is at the target (shown as bullseye); and (ii) the reactant is in the reactive state.}
\end{centering}
\end{figure}

The framework developed herein is also used to unravel the existence of universal features of gated reactions. Specifically, we point to the effect of fluctuations in the time between consecutive molecular collisions and show that these always lead to an increase in the \emph{mean} completion time of a gated reaction. Moreover, in cases where wild fluctuations lead to diverging means, we show that the exponent governing the heavy-tailed asymptotics of the ungated reaction is universally induced upon the gated reaction; thus generalizing earlier observations made for simple diffusion \cite{budde1995transient,caceres1995theory,mercado2019first}. Finally, the framework is utilized to expose exotic kinetic features that arise due to gating. As we show, these can be observed even in simple model systems such as the 1d random walk, yet they have so far been overlooked. Recent advancements in single-molecule technologies make us hopeful that predictions coming from our framework will soon be tested experimentally.

\textit{Gated reactions on networks.}---The set of problems we analyze below is formulated as follows: Consider a reactant (particle) performing a continuous-time random walk (CTRW) \cite{CTRW1} on a network (see Fig. 1, for an example of a finite network embedded in a 2D lattice). In addition to its spatial motion, the reactant also undergoes stochastic internal dynamics flipping between a reactive state and non-reactive state. A reaction is deemed to occur if the reactant is found in the reactive state while its spatial position overlaps with the origin --- a designated target site which, e.g., harbours a complementary reactant. To keep things general, we make no assumptions on the network on top of which the dynamics takes place, the distributions that govern waiting times in the different sites, and the distributions that govern the random motion (jumps) of the particle between network sites. 

To progress, we assume that the position of the particle is decoupled from its internal state. The strategy is then to ``divide and rule". First, we tackle the reactant's internal dynamics which is hereby assumed to be governed by a continuous-time Markov chain composed of two states: Reactive (R) and Non-Reactive (NR). The transition rate from R to NR is denoted by $\alpha$ and the transition rate from NR to R is denoted by $\beta$ (Fig. 1, top-left). We are interested in the conditional probability to be in each of the states after time $t$, given  an initial internal state $q_{0} \in \{\text{R, NR}\}$. For example, when $q_{0}=\text{NR}$, this is given by $\text{P}(\text{R},t \mid \text{NR}) =\pi_{\textrm{R}}(1-e^{-\lambda t})$ and $\text{P}(\text{NR},t \mid \text{NR}) =\pi_{\textrm{NR}}+\pi_{\textrm{R}}e^{-\lambda t}$, where $\lambda=\alpha+\beta$ is the effective relaxation rate and with $\pi_{\textrm{R}}=\beta /\lambda$ and $\pi_{\textrm{NR}}=\alpha /\lambda$ standing for the equilibrium occupancies.

Consider now the time $T(\Vec{x}_{0})$ it takes the particle to react given an initial location $\Vec{r}_{0}$ and an initial internal state $q_{0}$ which we jointly denote by $\Vec{x}_{0}=(\Vec{r}_{0},q_{0})$.  Two contributions feed into $T(\Vec{x}_{0})$ as illustrated in Fig. 2: (i) $T_{FP}(\Vec{r}_{0})$ which is the first-passage time of the particle to the origin; and (ii) the time it takes the particle to react after it has first reached the origin. If the particle arrives at the origin in the reactive state it reacts immediately. Otherwise, the particle---which is now found at the origin in the non-reactive state---can be seen to start its motion anew with the following initial conditions $ \Vec{0}_{\text{NR}} \equiv(\Vec{0},\text{NR}$). Thus, letting $T(\Vec{0}_{\text{NR}})$ denote the  reaction time starting from $\Vec{0}_{\text{NR}}$, we have  
\begin{equation}  \label{eq:1}
T(\Vec{x}_{0}) = T_{FP}(\Vec{r}_{0}) + I_{FP} T(\Vec{0}_{\text{NR}}),
\end{equation} 
\noindent where $I_{FP}$ is an indicator random variable that receives the value 1 if the particle first arrived at the origin in the non-reactive state and 0 otherwise. 

Taking expectations in Eq. (\ref{eq:1}) we find \cite{SM}
\begin{equation}  \label{eq:2}
\Braket{T(\Vec{x}_{0})} = \Braket{T_{FP}(\Vec{r}_{0})} +\Big[ \pi_{NR} \pm (1-\pi_{q_{0}})\tilde{T}_{FP}(\Vec{r}_{0},\lambda) \Big]\Braket{T(\Vec{0}_\text{NR})},
\end{equation}
\noindent where we have a plus sign if $q_{0}=\text{NR}$, and a minus sign if $q_{0}=\text{R}$, and where $\tilde{T}_{FP}(\Vec{r_0},\lambda)=\langle e^{-\lambda T_{FP}(\Vec{r}_{0})}\rangle$ is the Laplace transform of $T_{FP}(\Vec{r}_{0})$ evaluated at $\lambda$. Note that in the limit $\lambda \Braket{T_{FP}(\Vec{r}_{0})} \gg 1$, i.e., when $\Vec{r}_{0}$ is far enough such that the internal state equilibrates before the particle arrives at the origin, the Laplace transform is negligible and the mean reaction time in Eq. (\ref{eq:2}) becomes independent of the initial internal state. 

The distribution of the gated reaction time can also be computed. Taking the Laplace transform of Eq. (\ref{eq:1}), we find \cite{SM} 
\begin{equation} \label{eq:3}
\begin{array}{cc}
  \tilde{T}(\Vec{x}_{0},s)=
 \pi_{\text{R}}\tilde{T}_{FP}(\Vec{r}_{0},s) + \pi_{\text{NR}}\tilde{T}_{FP}(\Vec{r}_{0},s)\tilde{T}(\Vec{0}_\text{NR},s)
 \\
\pm (1-\pi_{q_{0}})\tilde{T}_{FP}(\Vec{r}_{0},s+\lambda)\Big[ \tilde{T}(\Vec{0}_\text{NR},s) -1 \Big],
\end{array}
\end{equation}
\noindent where we have a plus sign if $q_{0}=\text{NR}$, and a minus sign if $q_{0}=\text{R}$. Note that here too, the above expression simplifies considerably in the limit  $\lambda \Braket{T_{FP}(\Vec{r}_{0})} \gg 1$ where $\tilde{T}_{FP}(\Vec{r}_{0},s+\lambda)$ is negligible, and one is left with the first two terms which are independent of the initial internal state. 

\begin{figure}[t]
\includegraphics[width=1\linewidth]{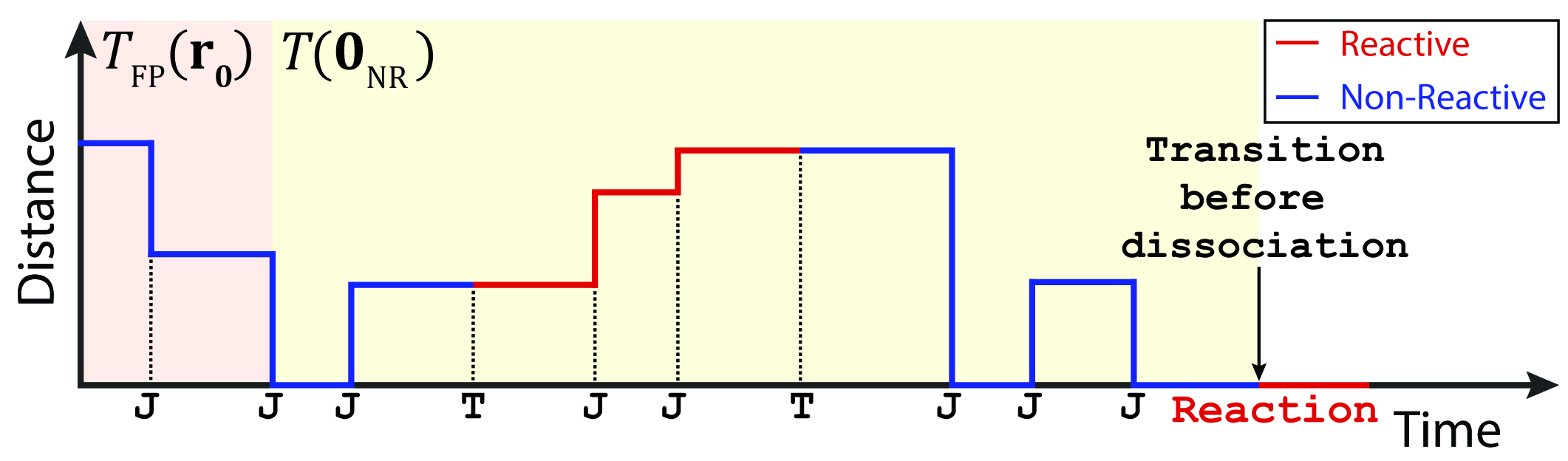}
\caption{The gated reaction time $T(\Vec{x}_{0})$ from Eq. (1) has two contributing factors: (i) the first-passage time of the particle to the origin $T_{FP}(\Vec{r}_{0})$; and (ii) the time it takes the particle to react after it has first reached the origin. In the illustration this time is given by $T(\Vec{0}_{\text{NR}})$, and $I_{FP}$ in Eq. (\ref{eq:1}) takes care of situations where the particle arrives at the origin in the reactive state and reacts immediately. On the time-axis, J stands for jump and T for transition between internal states. }
\end{figure}

Examining Eqs. (\ref{eq:2}) and (\ref{eq:3}) above reveals that part of the work required to determine the mean and distribution of a gated reaction time can be reduced to the solution of a standard (ungated) first passage time problem. Namely, one requires $T_{FP}(\Vec{r}_{0})$ which can be obtained by standard methods \cite{klafter2011first}. However, we are still left with the task of computing the gated reaction time $T(\Vec{0}_\text{NR})$. The mean and distribution of this random time will be our main focus going forward. 

\begin{figure*}[t]
\includegraphics[width=1\linewidth]{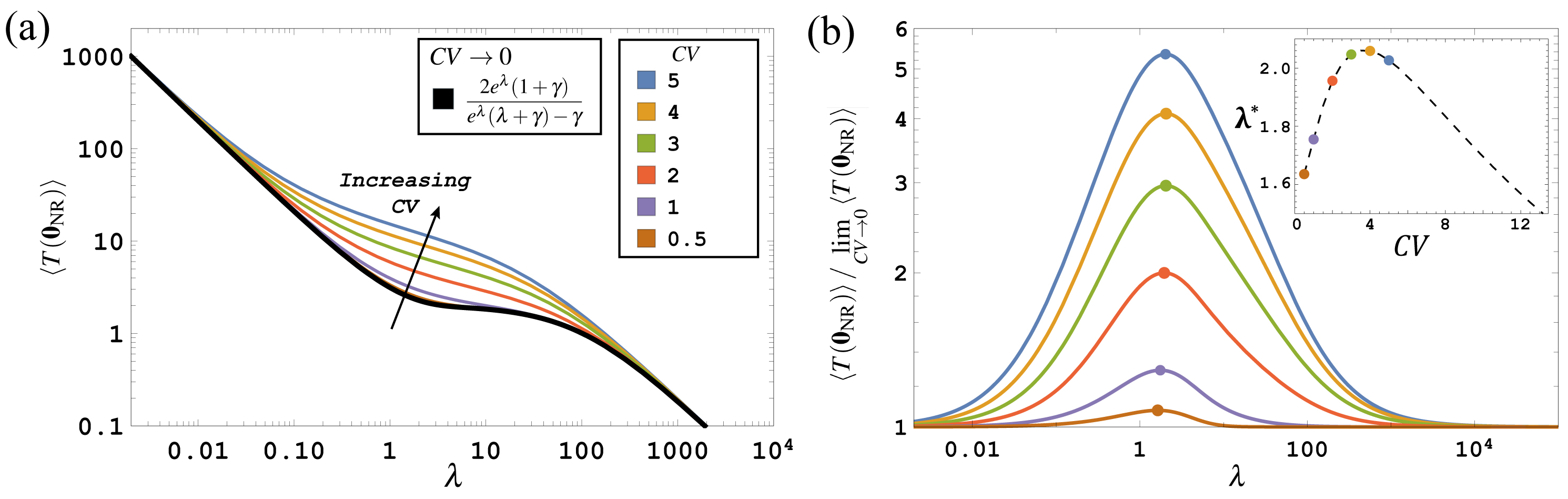}
\caption{Fluctuations in the return time to the origin increase the mean completion time of a gated reaction. Panel a). The mean reaction time $\braket{T(\Vec{0}_\text{NR})}$ from Eq. (\ref{eq:5}) vs. the internal relaxation rate $\lambda=\alpha+\beta$. Here we consider the symmetric case $\alpha=\beta=\lambda/2$ with $\gamma=100$, and  illustrate the effect by taking the return time $T_{FP}(\Vec{X_1})$ from a Gamma distribution with unit mean and increasing values of the coefficient of variation $CV$ (see \cite{SM} for details). The black line corresponds to the deterministic, $CV=0$, case which gives rise to the lower bound in Eq. (\ref{eq:6}). Higher $CV$s lead to higher mean completion times as predicted. It can be appreciated that the asymptotic $\braket{T(\Vec{0}_\text{NR})}\sim\beta^{-1}\sim\lambda^{-1}$ behaviour is common to all curves in both the high and low $\lambda$ regimes [see Eq. (\ref{eq:5}) and discussion below]. Panel b). The mean reaction time normalized by its lower bound: emphasizing that the behaviour at intermediate $\lambda$s is sensitive to the $CV$ of the return time to the origin. Furthermore, each curve has a unique value $\lambda^{*}$ for which the deviation from the lower bound is maximal. This value can be determined by solving a transcendental equation (inset, see \cite{SM} for details).} 
\end{figure*}

Let $W_{\Vec{0}}$ denote the random time the particle waits at the origin before jumping to a different site, and let $\Vec{X_1}$ stand for the random location of the particle following this jump. Setting $W_{NR}$ as the random waiting time of the particle in the non-reactive state, we observe that the gated reaction time $T(\Vec{0}_\text{NR})$ can be written as follows 
\begin{equation}  \label{eq:4}
T(\Vec{0}_\text{NR})=\left\{\begin{array}{ll}W_{NR} & \text { if } W_{NR}<W_{\Vec{0}}, \\ 
W_{\Vec{0}} +T_{FP}(\Vec{X_1}) +I_{FP} T^{\prime}(\Vec{0}_\text{NR}) & \text { if } W_{NR} \geq W_{\Vec{0}},\end{array}\right.
\end{equation}
\noindent where $T_{FP}(\Vec{X_1})$ is the first-passage time to the origin starting from $\Vec{X_1}$, $I_{FP}$ is defined as before, and  $T^{\prime}(\Vec{0}_\text{NR})$ is an IID copy of $T(\Vec{0}_\text{NR})$. Indeed, if $W_{NR}<W_{\Vec{0}}$, the particle transitions to the reactive state before jumping out of the origin and a reaction immediately follows. Conversely, if $W_{NR} \geq W_{\Vec{0}}$, the particle jumps out of the origin before transitioning to the reactive state. It then requires a time $T_{FP}(\Vec{X_1})$ to return to the origin, and an additional gated reaction time  $T^{\prime}(\Vec{0}_\text{NR})$ provided it returned in the non-reactive state. We thus see that the gated reaction time $T(\Vec{0}_\text{NR})$ can be expressed in terms of an ungated first-passage time $T_{FP}(\Vec{X_1})$ and the waiting times $W_{NR}$ and $W_{\Vec{0}}$.

\textit{The mean completion time of a gated reaction.}---To proceed, we recall that here we assumed Markovian internal dynamics which means that $W_{NR}$ is exponentially distributed with rate $\beta$. We will now also assume that $W_{\Vec{0}}$ is exponentially distributed with rate $\gamma$, but note that this assumption is made for clarity and that it can be easily generalized. Also, note that waiting times in other sites are kept general. Taking expectations in Eq. (\ref{eq:4}) we find \cite{SM}
\begin{equation} \label{eq:5}
 \braket{T(\Vec{0}_\text{NR})} =   \frac{\gamma^{-1}+\braket{T_{FP}(\Vec{X_1})}}{K_{D} + \pi_{\textrm{R}}\Big[1-\tilde{T}_{FP}(\Vec{X_1}, \lambda)\Big]},
\end{equation}
\noindent where $K_{D}=\beta/\gamma$, $\pi_{\textrm{R}}=\beta /\lambda$, and with $\lambda=\alpha+\beta$. Equation (\ref{eq:5}) asserts that $\braket{T(\Vec{0}_\text{NR})}$ can be determined provided the mean and distribution of the ungated return time $T_{FP}(\Vec{X_1})$.

To better understand Eq. (\ref{eq:5}), we observe that the $\braket{T(\Vec{0}_\text{NR})}\sim\beta^{-1}$ asymptotics holds for both large and small value of $\beta$. Specifically, when $\beta \gg 1$ a reaction is almost sure to happen before the particle leaves the origin. In this limit one can safely neglect $T(\Vec{0}_\text{NR})$ in Eqs. (\ref{eq:1}) and (\ref{eq:2}) concluding that the gated and ungated problems are practically equivalent. In the other extreme, $\beta \ll 1$, the transition to the non-reactive state becomes rate limiting and $\braket{T(\Vec{0}_\text{NR})}\gg1$. 

Next, consider the limit $\gamma \gg 1$ in which the particle leaves the origin almost instantaneously and without reacting. We then find $\braket{T(\Vec{0}_\text{NR})} = \pi_{\textrm{R}}^{-1}\Big[1-\tilde{T}_{FP}(\Vec{X_1}, \lambda)\Big]^{-1} \braket{T_{FP}(\Vec{X_1})}$. To understand this result observe that in this limit the probability that the particles returns to the origin in the reactive state is given by  $p=\pi_{\textrm{R}}\Big[1-\tilde{T}_{FP}(\Vec{X_1}, \lambda)\Big]$. Thus, on average, the particle returns to the origin $1/p$ times before a reaction occurs with each return taking $\braket{T_{FP}(\Vec{X_1})}$ time units on average. In the other extreme, $\gamma \ll 1$, the particle is slow to leave the origin and a reaction occurs following a transition to the non-reactive state. In this limit we find $\braket{T(\Vec{0}_\text{NR})}=1/\beta$. 

The mean reaction time in Eq. (\ref{eq:5}) scales linearly with the mean of $T_{FP}(\Vec{X_1})$ as expected. To better understand how fluctuations in the return time to the origin affect the result, we neglect the dissociation-time $\gamma^{-1}$ and expand  $\tilde{T}_{FP}(\Vec{X_1},\lambda)$ by its moments to second order in $\lambda$. Doing this we obtain $\braket{T(\Vec{0}_\text{NR})} \simeq \beta^{-1} + \frac{1}{2}\pi^{-1}_{\textrm{R}}(1+CV^2) \braket{T_{FP}(\Vec{X_1})}$, where $CV=\sigma(T_{FP}(\Vec{X_1}))/\braket{T_{FP}(\Vec{X_1})}$ stands for the coefficient of variation \cite{SM}. We thus see that fluctuations in $T_{FP}(\Vec{X_1})$ tend to increase the mean completion time of the gated reaction. In fact, invoking Jensen's inequality we get 
\begin{equation} \label{eq:6}
 \braket{T(\Vec{0}_\text{NR})} \geq \frac{\gamma^{-1}+\braket{T_{FP}(\Vec{X_1})}}{K_{D} + \pi_{\textrm{R}}\Big[1-e^{-\lambda \braket{T_{FP}(\Vec{X_1})}}\Big]}
\end{equation}
\noindent where the equality holds if and only if $\sigma(T_{FP}(\Vec{X_1}))=0$. A  deterministic return time thus yields a universal lower bound for $\braket{T(\Vec{0}_\text{NR})}$, and any fluctuation around the mean return time $\braket{T_{FP}(\Vec{X_1})}$ will necessarily slow down the completion of the gated reaction. This effect is illustrated in Fig. 3.

\textit{Beyond the mean.}---We now turn attention to the distribution of $T(\Vec{0}_\text{NR})$. Laplace transforming Eq. (\ref{eq:4}) we get \cite{SM}
\begin{equation} \label{eq:7}
\tilde{T}(\Vec{0}_\text{NR},s)=
\frac{\pi^{-1}_{\textrm{R}}K_{D} + \tilde{T}_{FP}(\Vec{X_1},s) - \tilde{T}_{FP}(\Vec{X_1},s + \lambda) }{\pi^{-1}_{\textrm{R}}(\frac{s}{\gamma}+K_{D}+1)-K_{eq}\tilde{T}_{FP}(\Vec{X_1},s)-\tilde{T}_{FP}(\Vec{X_1},s + \lambda)},  
\end{equation}
\noindent where $K_{eq}=\alpha/\beta$; and it can once again be appreciated that the gated reaction time can be put in terms of the ungated return time to the origin $T_{FP}(\Vec{X_1})$.

Equation (\ref{eq:7}) is indispensable in cases where $\braket{T_{FP}(\Vec{X_1})}$ diverges. Equation (\ref{eq:5}) then provides little information, but Eq. (\ref{eq:7}) can still be used to e.g., show that  ${T(\Vec{0}_\text{NR})}$ inherits the tail asymptotics of $T_{FP}(\Vec{X_1})$. Specifically we find that if  $\tilde{T}_{FP}(\Vec{X_1},s)\simeq1-(\tau s)^{\theta}$ for $s\ll1$, with $0<\theta<1$ and $\tau>0$, then \cite{SM}
\begin{equation}\label{eq:8}
\tilde{T}(\Vec{0}_\text{NR},s) \simeq 1 - \frac{\pi^{-1}_{\textrm{R}} }{1-\tilde{T}_{FP}(\Vec{X_1}, \lambda)+\lambda/\gamma} (\tau s)^{\theta}.  
\end{equation}
\noindent From here it follows that if the survival function of $T_{FP}(\Vec{X_1})$ decays as $\sim t^{-\theta}$, then so does that of ${T(\Vec{0}_\text{NR})}$; and the prefactor can be determined by Eq. (\ref{eq:8}) and the Tauberian theorem. 

\textit{Putting it all to work.}---To illustrate the applicability of our approach, consider now a simple symmetric random walk on a 1d lattice. 
Starting the walk at $\Vec{r}_{0}$, and working in discrete time, it is well known that the Z-transform of the first-passage time to the origin is given by $\hat{T}_{FP}(\Vec{r}_{0},z)=(\frac{1 - \sqrt{1-z^2}}{z})^{\abs{\Vec{r}_{0}}}$ \cite{steele2012stochastic}. The corresponding solution in continuous time is then given by $\tilde{T}_{FP}(\Vec{r}_{0},s)=(\frac{1 - \sqrt{1-{\tilde{\psi}(s)}^2}}{\tilde{\psi}(s)})^{\abs{\Vec{r}_{0}}}$, where we have simply replaced $z$ with the Laplace transform of the waiting-time distribution, $\tilde{\psi}(s)$, in the CTRW \cite{bel2006random}. Finally, we trivially observe that when such a symmetric random walk leaves the origin it will be found at $\pm1$ with equal probability. As the return time from these lattice points to the origin is equal in law, we have   $\tilde{T}_{FP}(\Vec{X_1},s)=\tilde{T}_{FP}(+1,s)$. Substituting these results into Eq. (\ref{eq:7}) and then (\ref{eq:3}), the solution to the corresponding gated problem is readily obtained; and we have verified that this solution identifies with the solution obtained by Budde, Cáceres and Ré for a particle that is initially prepared in the reactive state \cite{budde1995transient,caceres1995theory}. In lieu of the general approach developed herein, the latter was obtained with admirable effort.

Continuing with the same example, we plot numerical inversions \cite{abate2004multi} of the reaction time distribution in Eq. (\ref{eq:3}) (Fig. 4). Two gated cases, and the corresponding ungated case, are compared. We first observe that the power-law governing the long time asymptotics of all distributions is identical to the $\sim t^{-3/2}$ decay which is characteristic to the first-passage probability of the 1d random walk. These asymptotic results agree with the prediction of Eq. (\ref{eq:8}). However, the behaviour at intermediate times differs significantly between gated cases. For the first case (orange), we take $\beta=1$ which gives an overall behaviour that is almost identical to that observed for the ungated case where the particle is always reactive. In the second case (red), we take $\beta=10^{-3}$ and find that this leads to the emergence of a wide intermediate time window that is governed by a $\sim t^{-1/2}$ power-law decay. This ``cryptic regime" was also observed in the analogous diffusion problem studied by  Mercado-Vásquez and Boyer \cite{mercado2019first}, but here we see that it can also be accompanied by an additional exotic and previously unobserved kinetic effect which renders the gate reaction time distribution multimodal (Fig. 4, inset). We trace this effect to the existence of two populations of particles: those which reached the origin without ever switching to the  non-reactive state and those which switched prior to reaching. As the latter are blocked from reacting for a mean time $1/\beta\gg1$ that is much larger than the median return time to the origin, two distinct peaks are formed. 

\begin{figure}[t]
\includegraphics[width=0.85\linewidth]{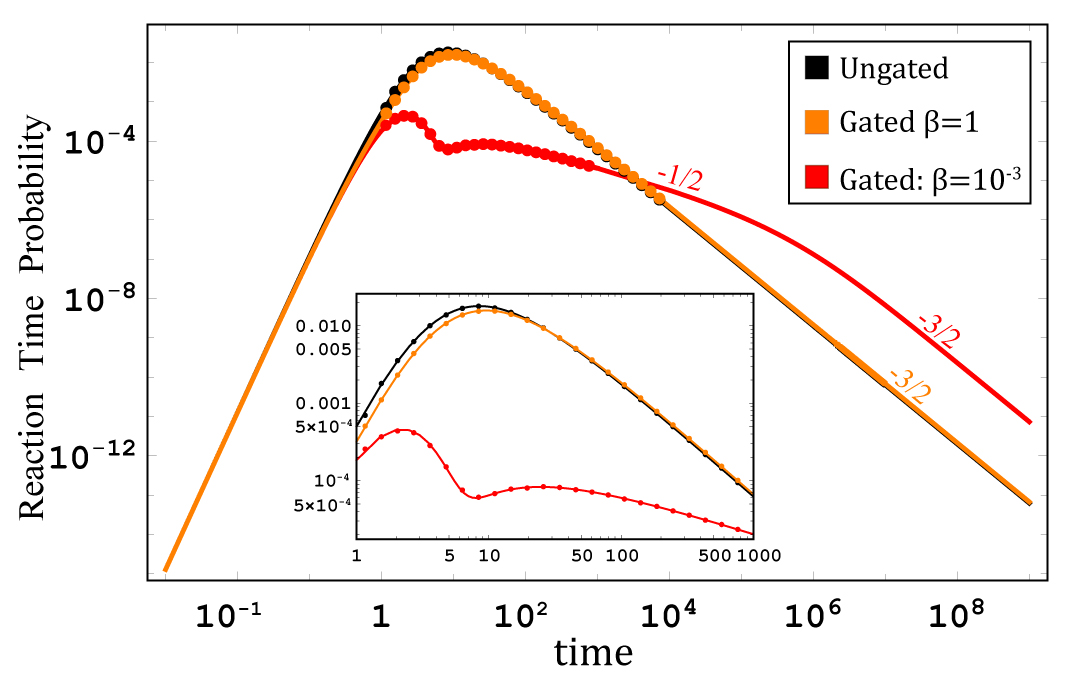}
\caption{Gated reaction time distributions for CTRW on a 1d lattice and the ungated behaviour for comparison. Here, the particle is taken to start from position $\Vec{r}_{0}=+5$ in the reactive state. The rate of transition to the non-reactive state is $\alpha=1$ and two different rates are considered for the reverse reaction: $\beta=10^{-3}$ (red) and $\beta=1$ (orange). The waiting time distribution in all sites (origin included) is taken to be exponential with rate $\gamma=1$. Solid lines come from numerical inversion of Eq. (\ref{eq:3}) and full circles from Monte-Carlo simulations.}
\end{figure}

\textit{Numerical applications.}---The approach taken for the 1d CTRW extends to many other first-passage time problems for which analytical results are known. In all such cases, the framework developed herein can be used to readily yield corresponding solutions for gated reaction times. However, in many complicated scenarios analytical results are not available and one must then resort to numerical simulations. We now explain how these can be considerably sped up by taking advantage of the relations derived above. Consider for example the mean time of a gated reaction, and imagine that the latter should be determined via numerical simulations for a wide range of rate constants $\alpha,\beta$ and $\gamma$. While this process can be extremely time consuming, one can instead take advantage of Eq. (\ref{eq:5}) which only requires numerical determination of the mean and distribution of the ungated first-passage time $T_{FP}(\Vec{X_1})$. As the latter does not depend on the above parameters it can be determined once and for all, thus providing a quick and efficient way of getting gated reaction times. Similar reasoning applies to the mean gated reaction time in Eq. (\ref{eq:2}), which in turn requires determination of one  additional ungated first-passage time: $T_{FP}(\Vec{r_0})$. In \cite{SM} we exemplify successful implementation of this procedure for the gated reaction depicted in Fig. 1.   

\textit{Conclusions.}---In this letter we developed a unified approach to gated reactions on networks, and the results obtained were used to extract considerable insight. Applications cover the entire spectrum of chemical reactions and are especially relevant to the emerging field of single-molecule chemistry. As our framework extends current knowledge on the long studied topic of first-passage it also applies more broadly, and shall be particularly useful in the context of search and foraging where similar ideas also apply.   

\textit{Acknowledgments.}---S.R. acknowledges support from the Azrieli Foundation, from the Raymond and Beverly Sackler Center for Computational Molecular and Materials Science at Tel Aviv University, and from the Israel Science Foundation (grant No. 394/19). The authors wish to thank Arnab Pal, Somrita Ray, Sarah Kostinski and Ofek Lauber Bonomo for fruitful conversations and advice.

{}
\end{document}





\setcounter{equation}{0}
\setcounter{figure}{0}
\setcounter{table}{0}
\setcounter{page}{1}
\setcounter{section}{0}
\makeatletter
\renewcommand{\theequation}{S\arabic{equation}}
\renewcommand{\thefigure}{S\arabic{figure}}
\renewcommand{\thesection}{S\Roman{section}} 
\renewcommand{\bibnumfmt}[1]{[S#1]}
\renewcommand{\citenumfont}[1]{S#1}

\begin{center}\Large{Supplemental Material for: ``A Unified Approach to Gated Reactions on Networks''}\end{center}

\author{{\normalsize{}Yuval Scher$^{1}$}
{\normalsize{}}}
\email{yuvalscher@mail.tau.ac.il}

\author{{\normalsize{}Shlomi Reuveni$^{1}$}
{\normalsize{}}}
\email{shlomire@tauex.tau.ac.il}

\affiliation{\noindent \textit{$^{1}$School of Chemistry, Center for the Physics \& Chemistry of Living Systems, Ratner Institute for Single Molecule Chemistry, and the Sackler Center for Computational Molecular \& Materials Science, Tel Aviv University, 6997801, Tel Aviv, Israel}}

\maketitle

\tableofcontents

\section{Derivation of Eq. (2) in the main text}

\noindent Taking expectations of both sides of Eq. (1) in the main text we get 
\begin{equation} 
\Braket{T(\Vec{x}_{0})} = \Braket{T_{FP}(\Vec{r}_{0})} + \Braket{I_{FP}} \Braket{T(\Vec{0}_{\text{NR}})} 
\end{equation}
\noindent where we have used the independence of $I_{FP}$ (the internal state at the moment of the first-passage time to the origin) and $T(\Vec{0}_{\text{NR}})$ (the gated reaction time when starting from  $\Vec{0}_{\text{NR}}\equiv(\Vec{0},\text{NR})$) to write: $\Braket{I_{FP}T(\Vec{0}_{\text{NR}})}=\Braket{I_{FP}} \Braket{T(\Vec{0}_{\text{NR}})}$. Indeed, the latter follows from the fact that the gated reaction time starting at the origin can only depend on $I_{FP}$ via the internal state, which in $T(\Vec{0}_{\text{NR}})$ has already been set to be non-reactive. We are then left with the task of calculating $\Braket{I_{FP}}$ which is given by
\begin{equation}
\Braket{I_{FP}} = 0 \cdot \Braket{\text{P}(\text{R},{T_{FP}(\Vec{r}_{0})} \mid q_{0})}   + 1 \cdot \Braket{\text{P}(\text{NR},{T_{FP}(\Vec{r}_{0})} \mid q_{0})}=\Braket{\text{P}(\text{NR},{T_{FP}(\Vec{r}_{0})} \mid q_{0})}  .    
\end{equation}
\noindent Explicitly, when $q_{0}=\text{NR}$, we have and $\text{P}(\text{NR},t \mid \text{NR}) =\pi_{\textrm{NR}}+\pi_{\textrm{R}}e^{-\lambda t}$, and so 
\begin{equation}
\Braket{\text{P}(\text{NR},{T_{FP}(\Vec{r}_{0})} \mid \textrm{NR})}=\pi_{\textrm{NR}}+\pi_{\textrm{R}}\tilde{T}_{FP}(\Vec{r}_{0},\lambda),
\end{equation}
\noindent where $\tilde{T}_{FP}(\Vec{r}_{0},\lambda)$ is the Laplace transform of $T_{FP}(\Vec{r}_{0})$ evaluated at $\lambda$.
Similarly, when $q_{0}=\text{R}$, we have $\text{P}(\text{NR},t \mid \text{R}) =\pi_{\textrm{NR}}(1-e^{-\lambda t})$, and so
\begin{equation}
\Braket{\text{P}(\text{NR},{T_{FP}(\Vec{r}_{0})} \mid \textrm{R})}=\pi_{\textrm{NR}}-\pi_{\textrm{NR}}\tilde{T}_{FP}(\Vec{r}_{0},\lambda).   
\end{equation}

\noindent Equation (2) in the main text follows directly from Eqs. (S1)-(S4) above. 

\section{Derivation of Eq. (3) in the main text}

\noindent Here again, we start from Eq. (1) in the main text. Laplace transforming it we get
\begin{equation} 
\tilde{T}(\Vec{x}_{0},s)=\Braket{e^{-sT(\Vec{x}_{0})}} = \Braket{e^{-s\Big[T_{FP}(\Vec{r}_{0}) +I_{FP} T(\Vec{0}_{\text{NR}})\Big]}},
\end{equation}
which in turn gives
\begin{equation}
\begin{array}{cc}
 \tilde{T}(\Vec{x}_{0},s)=\Braket{\text{P}(\text{R},T_{FP}(\Vec{r}_{0}) \mid q_{0})e^{-sT_{FP}(\Vec{r}_{0})}}+\Braket{\text{P}(\text{NR},T_{FP}(\Vec{r}_{0}) \mid q_{0}) e^{-s\Big[T_{FP}(\Vec{r}_{0}) + T(\Vec{0}_{\text{NR}})\Big]} }     
 \\
 =\Braket{\text{P}(\text{R},T_{FP}(\Vec{r}_{0}) \mid q_{0})e^{-sT_{FP}(\Vec{r}_{0})}}+\Braket{\text{P}(\text{NR},T_{FP}(\Vec{r}_{0}) \mid q_{0}) e^{-sT_{FP}(\Vec{r}_{0})} }  \Braket{e^{-s T(\Vec{0}_{\text{NR}})}}.      
\end{array}
\end{equation}

\noindent Setting $q_{0}=\textrm{NR}$ and using $\text{P}(\text{R},t \mid \text{NR}) =\pi_{\textrm{R}}(1-e^{-\lambda t})$ and $\text{P}(\text{NR},t \mid \text{NR}) =\pi_{\textrm{NR}}+\pi_{\textrm{R}}e^{-\lambda t}$, we get
\begin{equation}
\begin{array}{cc}
 \tilde{T}(\Vec{x}_{0},s)= \Braket{ \pi_{\text{R}}e^{-sT_{FP}(\Vec{r}_{0})}-\pi_{\text{R}}e^{-(s+\lambda)T_{FP}(\Vec{r}_{0})}}+\Braket{\pi_{\text{NR}}e^{-sT_{FP}(\Vec{r}_{0})}+\pi_{\text{R}} e^{-(s+\lambda)T_{FP}(\Vec{r}_{0})}}\Braket{e^{-s T(\Vec{0}_{\text{NR}})}}        
       \\
=\pi_{\text{R}}\tilde{T}_{FP}(\Vec{r}_{0},s) + \pi_{\text{NR}}\tilde{T}_{FP}(\Vec{r}_{0},s)\tilde{T}(\Vec{0}_\text{NR},s)
- \pi_{\text{R}}\tilde{T}_{FP}(\Vec{r}_{0},s+\lambda) + \pi_{\text{R}}\tilde{T}(\Vec{0}_\text{NR},s)  \tilde{T}_{FP}(\Vec{r}_{0},s+\lambda),
\end{array}      
\end{equation}

\noindent Similarly, setting $q_{0}=\textrm{R}$ and using $\text{P}(\text{R},t \mid \text{R}) =\pi_{\textrm{R}}+\pi_{\textrm{NR}}e^{-\lambda t}$ and $\text{P}(\text{NR},t \mid \text{R}) =\pi_{\textrm{NR}}\Big(1-e^{-\lambda t}\Big)$, we get
\begin{equation}
\begin{array}{cc}
 \tilde{T}(\Vec{x}_{0},s)= \Braket{ \pi_{\text{R}}e^{-sT_{FP}(\Vec{r}_{0})}+\pi_{\text{NR}}e^{-(s+\lambda)T_{FP}(\Vec{r}_{0})}}+\Braket{\pi_{\text{NR}}e^{-sT_{FP}(\Vec{r}_{0}) }-\pi_{\text{NR}} e^{-(s+\lambda)T_{FP}(\Vec{r}_{0})}}\Braket{e^{-s T(\Vec{0}_{\text{NR}})}}        
       \\
 =\pi_{\text{R}}\tilde{T}_{FP}(\Vec{r}_{0},s) + \pi_{\text{NR}}\tilde{T}_{FP}(\Vec{r}_{0},s)\tilde{T}(\Vec{0}_\text{NR},s)
+ \pi_{\text{NR}}\tilde{T}_{FP}(\Vec{r}_{0},s+\lambda) - \pi_{\text{NR}}\tilde{T}(\Vec{0}_\text{NR},s)  \tilde{T}_{FP}(\Vec{r}_{0},s+\lambda),

\end{array}      
\end{equation}

\noindent Equation (3) in the main text follows directly from Eqs. (S5)-(S8) above.

\section{Derivation of Eq. (5) in the main text}

\noindent To derive Eq. (5) in the main text, we start from Eq. (4) in the main text and recall that the internal dynamics was assumed to be Markovian. Thus, the probability density function of the waiting time in the non-reactive state is given by $\phi(t) = \beta e^{-\beta t}$, and the cumulative distribution function is correspondingly given by  $\Phi(t) = P(W_{NR}<t) = 1 - e^{-\beta t}$. To proceed, we also assume that the waiting time of the particle at the origin is taken from an exponential distribution with probability density function $\psi(t) = \gamma e^{-\gamma t}$, and cumulative distribution function $\Psi(t) = P(W_{0}<t) = 1 - e^{-\gamma t} $.

We now use Eq. (4) to write the left hand side of Eq. (5) as
\begin{equation} 
\Braket{T(\Vec{0}_{\text{NR}})} = P(W_{NR}<W_{\Vec{0}}) \Braket{W_{NR} |  W_{NR}<W_{\Vec{0}}} +
P(W_{NR} \geq W_{\Vec{0}}) \braket{W_{\Vec{0}} +T_{FP}(\Vec{X_1}) +I_{FP} T^{\prime}(\Vec{0}_\text{NR})  |  W_{NR} \geq W_{\Vec{0}}} := \circled{1} + \circled{2}.
\end{equation}
To obtain the desired result, we start by evaluating $\circled{1}$ as
\begin{equation} 
 \circled{1} = P(W_{NR}<W_{\Vec{0}}) \Braket{W_{NR} |  W_{NR}<W_{\Vec{0}}} =  P(W_{NR}<W_{\Vec{0}}) \frac{ \int_{0}^{\infty} t \phi(t) (1-\Psi(t))\,dt  } {P(W_{NR}<W_{\Vec{0}})} = \int_{0}^{\infty} t \phi(t) (1-\Psi(t))\,dt = \frac{\beta}{(\beta + \gamma)^2}.
\end{equation}

We then move on to calculate the second, more challenging, term $\circled{2}$. Utilizing the the fact that $T_{FP}(\Vec{X_1})$,  $I_{FP}$ and $T^{\prime}(\Vec{0}_\text{NR})$ are independent of $W_{0}$ and $W_{NR}$, we obtain
\begin{equation}
    \circled{2} = P(W_{NR} \geq W_{\Vec{0}})\braket{W_{\Vec{0}} |  W_{NR} \geq W_{\Vec{0}}} + P(W_{NR} \geq W_{\Vec{0}})\Big[\braket{T_{FP}(\Vec{X_1})} +\Braket{I_{FP} T^{\prime}(\Vec{0}_\text{NR} )}\Big] := \circled{A} + \circled{B}.
\end{equation}
\noindent Calculation of $\circled{A}$ is done in the same manner as in Eq. (S10) to give $\gamma/ (\beta + \gamma)^2$. For $\circled{B}$ we note that $P(W_{NR} \geq W_{\Vec{0}})=\gamma/(\beta+\gamma)$ and recall that  $\Braket{I_{FP} T^{\prime}(\Vec{0}_\text{NR} )}=\Braket{I_{FP}}\braket{ T^{\prime}(\Vec{0}_\text{NR} )}$ from independence. We then use the fact that this is a renewal process, hence $T(\Vec{0}_\text{NR})$ and $T^{\prime}(\Vec{0}_\text{NR})$ are IID and in particular: $\braket{T^{\prime}(\Vec{0}_\text{NR})}=\braket{ T(\Vec{0}_\text{NR} )}$. To get $\Braket{I_{FP}}$ we use Eq. (S3) where we recall that here  $q_{0}=\text{NR}$ by construction. Plugging everything back together and rearranging we obtain Eq. (5) in the main text
\begin{equation}
 \braket{T(\Vec{0}_\text{NR})} =   \frac{\frac{1}{\gamma}+\braket{T_{FP}(\Vec{X_1})}}{K_{D} + \pi_{\textrm{R}}\Big[1-\tilde{T}_{FP}(\Vec{X_1}, \lambda)\Big]},
\end{equation}
with $K_{D}=\frac{\beta}{\gamma}$.

\section{Small $\lambda$ expansion of  $\Braket{T(\Vec{0}_{\text{NR}})}$}

\noindent Taking the limit $\gamma \to \infty$ in Eq. (5), and expanding  $\tilde{T}_{FP}(\Vec{X_1},\lambda)$ by its moments to second order in $\lambda$, we get

\begin{equation}
\braket{T(\Vec{0}_\text{NR})}=\frac{\pi^{-1}_{\textrm{R}}\braket{T_{FP}(\Vec{X_1})}}{\lambda\braket{T_{FP}(\Vec{X_1})}-\frac{1}{2}\lambda^2\braket{T_{FP}(\Vec{X_1})^2}}.
\end{equation}

\noindent Rearranging,

\begin{equation}
\braket{T(\Vec{0}_\text{NR})}=\frac{\pi^{-1}_{\textrm{R}}}{\lambda}\frac{1}{1-\frac{1}{2}\lambda\frac{\braket{T_{FP}(\Vec{X_1})^2}}{\braket{T_{FP}(\Vec{X_1})}}}=\beta^{-1}\frac{1}{1-\frac{1}{2}\lambda\frac{\braket{T_{FP}(\Vec{X_1})^2}}{\braket{T_{FP}(\Vec{X_1})}}}.
\end{equation}

\noindent By definition of the variance $\sigma^2(T_{FP}(\Vec{X_1}))=\braket{T_{FP}(\Vec{X_1})^2}-\braket{T_{FP}(\Vec{X_1})}^2$, so we can rewrite Eq. (S14) as
\begin{equation}
\braket{T(\Vec{0}_\text{NR})}=\beta^{-1}\frac{1}{1-\frac{1}{2}\lambda\Big(\braket{T_{FP}(\Vec{X_1})}+\frac{\sigma^2}{\braket{T_{FP}(\Vec{X_1})}}\Big)}.\end{equation} 
\noindent Recalling that $\frac{1}{1+x}\approx1-x$ for $x\ll1$, we have
\begin{equation}
\braket{T(\Vec{0}_\text{NR})}=\beta^{-1} \Big[ 1 +  \frac{1}{2}\lambda\Big(\braket{T_{FP}(\Vec{X_1})}+\frac{\sigma^2}{\braket{T_{FP}(\Vec{X_1})}} +\dots \Big)   \Big] = \beta^{-1} + \frac{1}{2}\pi^{-1}_{\textrm{R}}(1+CV^2)\braket{T_{FP}(\Vec{X_1})} + \dots, 
\end{equation}

\noindent wherewe have used $\frac{\lambda}{\beta}=\pi^{-1}_{\textrm{R}}$ and the definition of the coefficient of variation $CV=\frac{\sigma(T_{FP}(\Vec{X_1}))}{\braket{T_{FP}(\Vec{X_1})}}$.


\section{Gamma-Distributed Return Times}

\noindent In Fig. 3 of the main text we consider a return time $T_{FP}(\Vec{X_1})$ that comes from a Gamma distribution with shape parameter $n$ and scale parameter $\theta$. The probability density function of this return time is then given by 
\begin{equation}
     f(t)=\frac{1}{\Gamma(n) \theta^{n}} t^{n-1} e^{-\frac{t}{\theta}},
\end{equation}
\noindent which gives a mean of $\braket{T_{FP}(\Vec{X_1})}=n\theta$, and the following Laplace transform 
\begin{equation}
     \braket{e^{-sT_{FP}(\Vec{X_1})}} = (1 + s\theta)^{-n}.
\end{equation}
\noindent Inserting these expressions into Eq. (5) in the main text, while setting $\alpha=\beta=\frac{1}{2}\lambda$, yields the mean gated reaction time 
\begin{equation}
    \braket{T(\Vec{0}_\text{NR})}=\frac{2(n \theta \gamma +1 )}{\lambda + \gamma -\gamma(1+ \theta \lambda)^{-n}}.
\end{equation}
To study the effect that fluctuations in the return time have on $\braket{T(\Vec{0})}$, we set $\theta=1/n$ such that the mean of $T_{FP}(\Vec{X_1})$ is set to 1 while its coefficient of variation is given by $CV=1/\sqrt{n}$. In this way, we can tune $CV$ by changing the shape parameter $n$, while keeping the mean of the return time distribution fixed.

In Fig. 3a of the mean text we plot the mean gated reaction time in Eq. (S19) for different values of $CV$. Note that the limit of $CV \to 0$, i.e., the case of deterministic return times, is obtained by taking $n \to \infty$, which gives
\begin{equation}
    \lim_{CV \to 0} \braket{T(\Vec{0}_\text{NR})}=\frac{2 e^{\lambda} (1 + \gamma)}{e^{\lambda}(\lambda + \gamma) - \gamma}.
\end{equation}
As we show in the main text, the expression in Eq. (S20) serves as a lower bound of the mean reaction time in Eq. (S19).

In Fig. 3b, we plot the mean reaction time normalized by its lower bound in Eq. (S20). We demonstrate that the deviation from the lower bound is CV dependent, and that for each CV value there is a corresponding $\lambda^{*}$ for which this deviation is maximal. To find this value we look for extrema of the normalized gated reaction time. We found that every extremum must satisfy the following transcendental equation:
\begin{equation}
e^{\lambda^{*}}=\frac{\left(n+\lambda^{*}\right)\left(1+\gamma+\lambda^{*}\right)\left(1+\frac{\lambda^{*}}{n}\right)^{n}-\gamma \lambda^{*}}{\lambda^{*}+n\left(1+\gamma+\lambda^{*}\right)},  
\end{equation}
\noindent which can be solved numerically to obtain $\lambda^{*}$ for any given value of $CV=1/\sqrt{n}$.

\section{Derivation of Eq. 7 in the main text}

\noindent To derive Eq. (7) in the main text, we start from Eq. (4) in the main text and once again recall that the internal dynamics was assumed to be Markovian. Thus, the probability density function of the waiting time in the non-reactive state is given by $\phi(t) = \beta e^{-\beta t}$, and the cumulative distribution function is correspondingly given by  $\Phi(t) = P(W_{NR}<t) = 1 - e^{-\beta t}$. To proceed, we also assume that the waiting time of the particle at the origin is taken from an exponential distribution with probability density function $\psi(t) = \gamma e^{-\gamma t}$, and cumulative distribution function $\Psi(t) = P(W_{0}<t) = 1 - e^{-\gamma t} $.

Laplace transforming Eq. 4 in the main text we obtain
\begin{equation} 
\tilde{T}(\Vec{0}_\text{NR},s)=\braket{e^{-sT(\Vec{0}_{\text{NR}})}} = P(W_{NR}<W_{\Vec{0}}) \braket{e^{-sW_{NR}} |  W_{NR}<W_{\Vec{0}}} +
P(W_{NR} \geq W_{\Vec{0}}) \braket{e^{-s\Big[W_{\Vec{0}} +T_{FP}(\Vec{X_1}) +I_{FP}T^{\prime}(\Vec{0}_{\text{NR}}) \Big]} |  W_{NR} \geq W_{\Vec{0}}} := \circled{1} + \circled{2}.
\end{equation}

\noindent We start by evaluating $\circled{1}$, which can be written explicitly as
\begin{equation} 
 \circled{1} = P(W_{NR}<W_{\Vec{0}}) \braket{e^{-sW_{NR}} |  W_{NR}<W_{\Vec{0}}} =  P(W_{NR}<W_{\Vec{0}}) \frac{\mathscr{L} \{\phi(t) (1-\Psi(t)) \}  } {P(W_{NR}<W_{\Vec{0}})} = \mathscr{L} \{\phi(t) (1-\Psi(t)) \} = \frac{\beta}{s+\beta + \gamma}.
\end{equation}

We then move on to calculate the second, more challenging, term $\circled{2}$. Utilizing the the fact that $T_{FP}(\Vec{X_1})$,  $I_{FP}$ and $T^{\prime}(\Vec{0}_\text{NR})$ are independent of $W_{0}$ and $W_{NR}$, we obtain
\begin{equation}
    \circled{2} = P(W_{NR} \geq W_{\Vec{0}}) \braket{e^{-sW_{\Vec{0}}} |  W_{NR} \geq W_{\Vec{0}}} \braket{e^{-s\Big[ T_{FP}(\Vec{X_1}) +I_{FP}T^{\prime}(\Vec{0}_{\text{NR}})\Big]}}.
\end{equation}
\noindent The first two terms being multiplied can be calculated in the same manner as in Eq. (S23) above to give $\mathscr{L} \{\psi(t) (1-\Phi(t)) \}=\frac{\gamma}{s+\beta + \gamma}$. The third term can be calculated as we did in Eq. (S6), to give
\begin{equation}
  \circled{2} =   \mathscr{L} \{\psi(t) (1-\Phi(t)) \}  \Big[ \braket{\text{P}(\text{R}, T_{FP}(\Vec{X_1}) \mid \text{NR})e^{-s T_{FP}(\Vec{X_1})}}
 +\braket{\text{P}(\text{NR}, T_{FP}(\Vec{X_1})\mid \text{NR})e^{-s T_{FP}(\Vec{X_1})}} \braket{e^{T^{\prime}(\Vec{0}_{\text{NR}})}}\Big].
\end{equation}
\noindent To evaluate the above we recall that $\text{P}(\text{R},t \mid \text{NR}) =\pi_{\textrm{R}}(1-e^{-\lambda t})$ and $\text{P}(\text{NR},t \mid \text{NR}) =\pi_{\textrm{NR}}+\pi_{\textrm{R}}e^{-\lambda t}$; and further use the fact that $T^{\prime}(\Vec{0}_{\text{NR}})$ is an IID copy of $T(\Vec{0}_{\text{NR}})$. The overall expression for the Laplace transform of the gated reaction time $T(\Vec{0}_\text{NR})$ can then be written as
\begin{equation}
\begin{array}{cc}
  \tilde{T}(\Vec{0}_\text{NR},s)=\braket{e^{-sT(\Vec{0}_{\text{NR}})}} = \frac{\beta}{s+\beta +\gamma} + \frac{\gamma}{s+\beta +\gamma} \Big[\pi_{\textrm{R}}\braket{e^{-sT_{FP}(\Vec{X_1})}} - \pi_{\textrm{R}} \braket{e^{-(s+\lambda)T_{FP}(\Vec{X_1})}}\Big]      \\
    +  \frac{\gamma}{s+\beta +\gamma}   \Big[ \pi_{\textrm{NR}}\braket{e^{-sT_{FP}(\Vec{X_1})}} + \pi_{\textrm{R}} \braket{e^{-(s+\lambda)T_{FP}(\Vec{X_1})}}  \Big]\braket{e^{-sT(\Vec{0}_{\text{NR}})}}.
\end{array}
\end{equation}

\noindent Rearranging to solve for $\braket{e^{-sT(\Vec{0})}}$, we obtain
\begin{equation}
 \tilde{T}(\Vec{0}_\text{NR},s)=\braket{e^{-sT(\Vec{0})}} = \frac{\frac{\beta}{s+\beta +\gamma} + \frac{\gamma}{s+\beta +\gamma} \Big[\pi_{\textrm{R}}\braket{e^{-sT_{FP}(\Vec{X_1})}} - \pi_{\textrm{R}} \braket{e^{-(s+\lambda)T_{FP}(\Vec{X_1})}}\Big]}{1-\frac{\gamma}{s+\beta +\gamma}   \Big[ \pi_{\textrm{NR}}\braket{e^{-sT_{FP}(\Vec{X_1})}} + \pi_{\textrm{R}} \braket{e^{-(s+\lambda)T_{FP}(\Vec{X_1})}} \Big]}.   
\end{equation}

\noindent  Multiplying both numerator and denominator by $\frac{s+\beta+\gamma}{ \pi_{\textrm{R}} \gamma}$, we get Eq. (7) in the main text
\begin{equation}
 \tilde{T}(\Vec{0}_\text{NR},s) = \frac{\frac{\beta}{\pi_{\textrm{R}} \gamma} + \braket{e^{-sT_{FP}(\Vec{X_1})}} -  \braket{e^{-(s+\lambda)T_{FP}(\Vec{X_1})}}}{\frac{s+\beta+\gamma}{ \pi_{\textrm{R}} \gamma}-K_{eq} \braket{e^{-sT_{FP}(\Vec{X_1})}} - \braket{e^{-(s+\lambda)T_{FP}(\Vec{X_1})}} }  := \frac{\pi^{-1}_{\textrm{R}}K_{D} + \braket{e^{-sT_{FP}(\Vec{X_1})}} -  \braket{e^{-(s+\lambda)T_{FP}(\Vec{X_1})}}}{\pi^{-1}_{\textrm{R}}(\frac{s}{\gamma}+K_{D} +1)-K_{eq} \braket{e^{-sT_{FP}(\Vec{X_1})}} - \braket{e^{-(s+\lambda)T_{FP}(\Vec{X_1})}} }.
\end{equation}

\section{Derivation of Eq. 8 in the main text}

\noindent To derive Eq. (8) in the main text we plug into Eq. (7) 
$\tilde{T}_{FP}(\Vec{X_1},s)\simeq1-(\tau s)^{\theta}$ for $s\ll1$, with $0<\theta<1$ and $\tau>0$. This gives
\begin{equation} 
\tilde{T}(\Vec{0}_\text{NR},s)\simeq
\frac{\pi^{-1}_{\textrm{R}}K_{D} + \Big[ 1-(\tau s)^{\theta} \Big] - \tilde{T}_{FP}(\Vec{X_1},s + \lambda) }{\pi^{-1}_{\textrm{R}}(\frac{s}{\gamma}+K_{D}+1)-K_{eq}\Big[ 1-(\tau s)^{\theta} \Big]-\tilde{T}_{FP}(\Vec{X_1},s + \lambda)},  
\end{equation}
\noindent which can be simplified by algebraic manipulations as follows 

\begin{equation} 
\tilde{T}(\Vec{0}_\text{NR},s)\simeq
\frac{\Big[\pi^{-1}_{\textrm{R}}K_{D} + 1-\tilde{T}_{FP}(\Vec{X_1},s + \lambda) \Big] \Big[1-\frac{(\tau s)^{\theta}}{\pi^{-1}_{\textrm{R}}K_{D} + 1-\tilde{T}_{FP}(\Vec{X_1},s + \lambda)]}\Big]  }{\Big[\pi^{-1}_{\textrm{R}}(K_{D}+1)-K_{eq}-\tilde{T}_{FP}(\Vec{X_1},s + \lambda)\Big]\Big[1+\frac{K_{eq}(\tau s)^{\theta}+\pi^{-1}_{\textrm{R}}\frac{s}{\gamma}}{\pi^{-1}_{\textrm{R}}(K_{D}+1)-K_{eq}-\tilde{T}_{FP}(\Vec{X_1},s + \lambda) } \Big]}.  
\end{equation}

\noindent Noting that $\pi^{-1}_{\textrm{R}}-K_{eq}=\frac{\alpha+\beta}{\beta}-\frac{\alpha}{\beta}=1$, we cancel matching terms and in numerator and denominator to obtain 

\begin{equation} 
\tilde{T}(\Vec{0}_\text{NR},s)\simeq
\Big[1 - \frac{(\tau s)^\theta}{\pi^{-1}_{\textrm{R}}K_{D} + 1 -  \tilde{T}_{FP}(\Vec{X_1},s + \lambda) }\Big] \frac{1}{1 + \frac{K_{eq}(\tau s)^\theta + \pi^{-1}_{\textrm{R}} \frac{s}{\gamma} }{\pi^{-1}_{\textrm{R}} (K_{D}+1) -K_{eq}-\tilde{T}_{FP}(\Vec{X_1},s + \lambda) }}.  
\end{equation}

\noindent In the limit $s \to 0$ we can neglect $\pi^{-1}_{\textrm{R}} \frac{s}{\gamma}$ with respect to $K_{eq}(\tau s)^\theta$, and further approximate $\tilde{T}_{FP}(\Vec{X_1},s + \lambda)\simeq\tilde{T}_{FP}(\Vec{X_1},\lambda)$. Then, expanding the fraction on the right, we obtain
\begin{equation} 
\tilde{T}(\Vec{0}_\text{NR},s)\simeq
\Big[1 - \frac{(\tau s)^\theta}{\pi^{-1}_{\textrm{R}}K_{D} + 1 - \tilde{T}_{FP}(\Vec{X_1},\lambda) }\Big] \Big[ 1 -  \frac{K_{eq}(\tau s)^\theta }{\pi^{-1}_{\textrm{R}} (K_{D}+1) -K_{eq}-\tilde{T}_{FP}(\Vec{X_1}, \lambda)}\Big].  
\end{equation}

\noindent Now by carefully multiplying these terms, neglecting higher order products and simplifying the result, we get Eq. (8) of the main text. 

\section{Analysing the gated reaction depicted in Fig. 1 of the main text}

\begin{figure}[h!]
\begin{centering}
\includegraphics[width=0.5\linewidth]{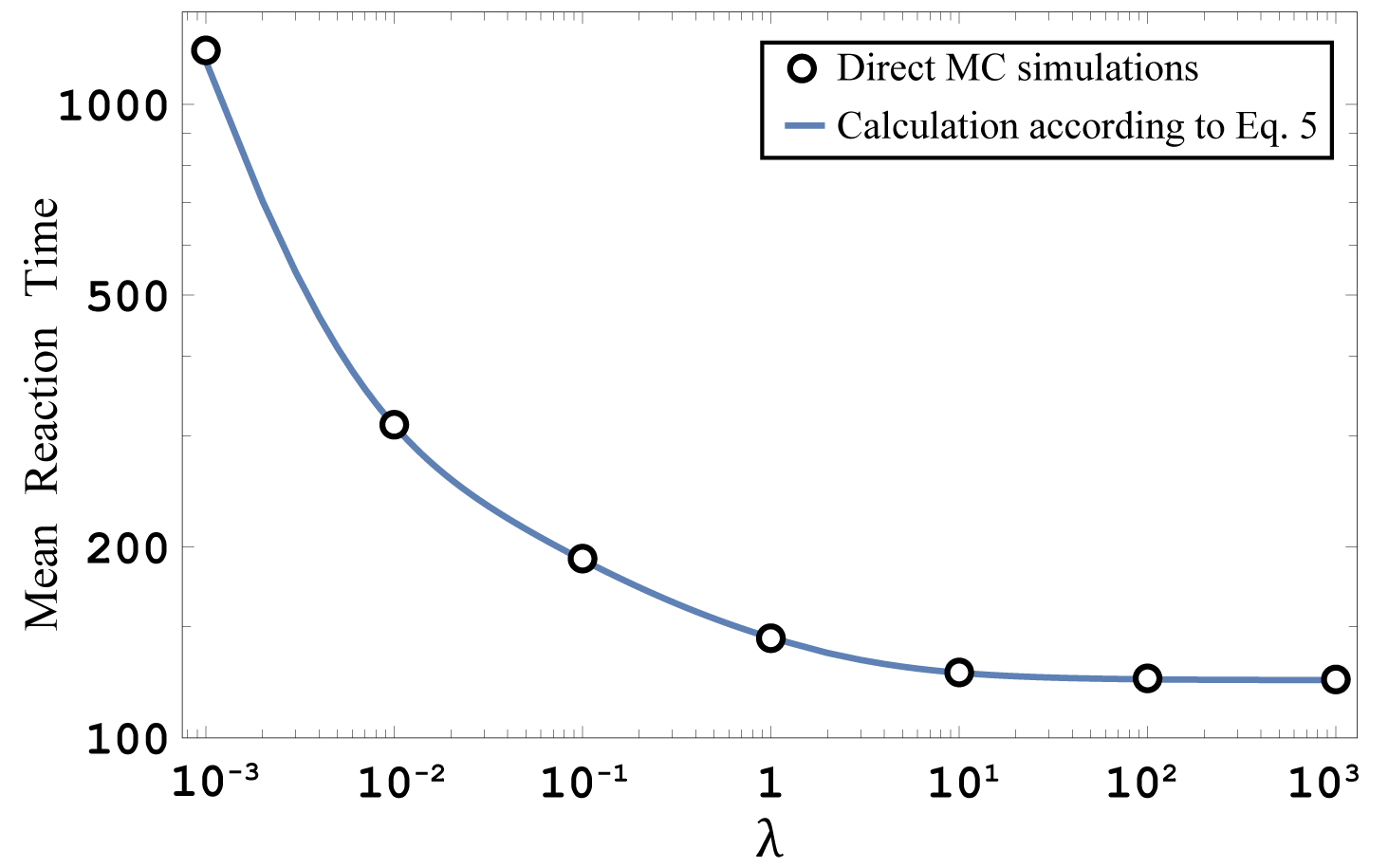}
\caption{The mean completion time of the gated reaction depicted in Fig. 1 of the main text vs $\lambda$. Results from direct Monte-Carlo simulations for a handful of $\lambda$ values (circles) are compared to the hybrid method in which Monte-Carlo simulations are only used to evaluate the distributions of the ungated first-passage times $T_{FP}(\Vec{X_1})$ and $T_{FP}(\Vec{r_0})$. These are then numerically Laplace transformed and fed into Eq. (5) which consequently feeds into Eq. (2) to determine the mean gated reaction time for all values of $\lambda$ (blue line). The hybrid method opens the door for exhaustive scan of governing parameters in complex gated reactions. Such exhaustive scans are virtually impossible with direct Monto-Carlo simulations as these are much slower in comparison.}
\end{centering}
\end{figure}

Lastly, we demonstrate how the framework developed in our letter can be used to speed up and greatly facilitate numerical studies of gated reactions on networks. To this end, we consider the gated reaction depicted in Fig. 1 of the main text. Taking the same spatial initial condition described in the figure, we set equilibrium initial conditions with $\pi_{NR}=\pi_{R}=1/2$ for the internal state (note that this implies $\lambda=2\alpha=2\beta$). For the waiting time of the particle in the different sites we have taken the exponential distribution with mean $1$, and further assumed that the particle jumps only to nearest neighbours and with equal probabilities. The mean reaction time was then obtained for a handful of $\lambda$ values by direct Monte-Carlo simulations (Fig. S1, circles). Note that for each value of $\lambda$ we performed $10^5$ simulations which serves to illustrate that exhaustive scan of the parameters space is impractical and computationally costly. In contrast, the mean reaction time can be readily computed using the hybrid method that we described in the main text. In this method, Monte-Carlo simulations are only used twice to estimate the ungated reaction times $T_{FP}(\Vec{X_1})$ and $T_{FP}(\Vec{r_0})$. This information is then used as input which feeds into Eqs. (5) and (2) to determine the mean completion time of the gated reaction for all values of $\lambda$ (Fig. S1, blue line).